\documentclass[twocolumn,amsmath,amssymb,pre,showpacs]{revtex4}

\usepackage{graphicx}
\usepackage{dcolumn}
\usepackage{bm}
\usepackage{amssymb}
\usepackage{amsmath}
\usepackage{amsthm,amsfonts}

\begin{document}
\bibliographystyle{aip}

\title{Heat, Work and Energy Currents in the Boundary-Driven XXZ Spin Chain}

\author{Emmanuel Pereira}
 \email{emmanuel@fisica.ufmg.br}
\affiliation{Departamento de F\'{\i}sica--Instituto de Ci\^encias Exatas, Universidade Federal de Minas Gerais, CP 702,
30.161-970 Belo Horizonte MG, Brazil}

\date{\today}

\begin{abstract}
We address the detailed study of the energy current and its components, heat and work, in the boundary-driven 1D $XXZ$ quantum model. We carry out the investigation by considering two different approaches present in the
literature. First, we take the repeated interaction scheme and derive the expressions for the currents of heat and work, exchanged between system and baths. Then,  we perform the derivation  of the energy current
 by means of a Lindblad master equation together with a  continuity equation, another approach which is recurrently used. A comparison between the obtained expressions allows us to show the consistency of both approaches, and, in the latter expression derived from the
  Lindblad equation, it allows us to split the energy, which comes from the
 baths to the system, into heat and work. The recognition of work in the process, that is recurrently ignored in studies of transport, enables us to understand thermodynamical aspects and to solve some imbroglios in the physics behind the energy current in the $XXZ$ spin chain.

\end{abstract}


\def \Z {\mathbb{Z}}
\def \R {\mathbb{R}}
\def \La {\Lambda}
\def \la {\lambda}
\def \ck {l}
\def \F {\mathcal{F}}
\def \M {\mathcal{M}}
\newcommand {\md} [1] {\mid\!#1\!\mid}
\newcommand {\be} {\begin{equation}}
\newcommand {\ee} {\end{equation}}
\newcommand {\ben} {\begin{equation*}}
\newcommand {\een} {\end{equation*}}
\newcommand {\bg} {\begin{gather}}
\newcommand {\eg} {\end{gather}}
\newcommand {\ba} {\begin{align}}
\newcommand {\ea} {\end{align}}
\newcommand {\tit} [1] {``#1''}


\maketitle

\let\a=\alpha \let\b=\beta \let\d=\delta \let\e=\varepsilon
\let\f=\varphi \let\g=\gamma \let\h=\eta    \let\k=\kappa \let\l=\lambda
\let\m=\mu \let\n=\nu \let\o=\omega    \let\p=\pi \let\ph=\varphi
\let\r=\rho \let\s=\sigma \let\t=\tau \let\th=\vartheta
\let\y=\upsilon \let\x=\xi \let\z=\zeta
\let\D=\Delta \let\G=\Gamma \let\L=\Lambda \let\Th=\Theta
\let\P=\Pi \let\Ps=\Psi \let\Si=\Sigma \let\X=\Xi
\let\Y=\Upsilon

\section{Introduction}

 The comprehension of the transport laws of matter and/or energy between systems and environments is a central issue of nonequilibrium statistical physics. Research branches, such as Phononics, Photonics, Spintronics,
Electronics, etc., are dedicated to the theoretical investigation as well as to the experimental manipulation of the different forms of transport. Phononics, in particular, is an old, fundamental and active branch, devoted to the study and control of the heat current. In the last decade, inspired by the success of modern electronics, a significant effort has been spent to build the phononic analogs of the electronic devices in order to manipulate the heat flow: thermal diodes, thermal transistors, thermal memories and other tools have been proposed and detailed analyzed \cite{BLiRMP}. Since Debye, the typical models for the
investigation of heat conduction in insulating solids are given by anharmonic classical chains of oscillators \cite{LLP, Dhar}, and so, most of these proposals of phononic devices involve these classical systems.  Consequently, in spite of the existence of several interesting results, there is a demand for different approaches able to treat quantum effects.
Thus, with such a motivation, as well as with the impulse given by  the advance of lithography and the resulting ambient of miniaturization,
 several recent works are oriented to the study of the energy transport in genuine quantum models \cite{Wu, Mendoza-A,  SPL}.

Actually, it is important to note that the investigation of the energy transport properties in low-dimensional quantum models is also enhanced  by several other
issues beyond these fundamental questions of Phononics: a good example is the interest in the properties of cold atoms and related phenomena, stimulated by the technological advances allowing for their manipulation
\cite{cold1, cold2}. The research on the theme is also stimulated by  many other problems directly related to understanding the features of the energy current in quantum spin chains. For example,  important questions involving the occurrence of
different regimes of transport: ballistic, diffusive, etc. \cite{MHWG}, in a scenario with several results and discordances \cite{ZX, Wu}. The role played by the presence of dephasing processes in the bulk of the system and the
possibility of current enhancement \cite{Mendoza-A}. Changes in the energy transport near to many-body localization-delocalization transitions \cite{Varna}. The role played on the energy current by symmetries in the associated master equation \cite{PopLi, Prapid2017}. And, as an important problem of the previous recalled Phononics, but now involving genuine quantum systems, we have researches aiming to build
 devices such as a quantum thermal transistor  \cite{Jou}.

An exhaustively investigated quantum model is the 1D $XXZ$ chain \cite{Mendoza-A, SPL, MHWG, xxz1, xxz2, xxz3, ZM, PS, LOK},  with interest in several different areas, such as nonequilibrium statistical physics, condensed-matter, optics, quantum information, etc.
An important version of this $XXZ$ spin chain is the boundary driven model, i.e., a chain with target polarization at the boundaries. Another relevant version is the case in which the chain is passively and weakly
coupled to thermal baths \cite{BP, GZ}. Considering thermodynamic issues, precisely, the interpretation of the energy current, there is a significant difference between the two versions. In the weak coupling case, for the usual
situation of a time independent Hamiltonian, no work can be performed on the system, and so, the energy current is given only by the heat which is exchanged with the baths. But, for the boundary-driven case, i.e., for chains with
target polarization at the edges, we do not have a free-work process (even for a time-independent Hamiltonian): part of the energy current is due to the work which, say, the driven-boundary operation brings to the system \cite{FBarra}.
Such a distinction between  heat and power in the energy current is ignored in several works in the literature of these quantum spin systems, but it is necessary to identify these two fundamental objects of thermodynamics in order to understand some transport phenomena, as we
 will make clear in the present paper.

It is pertinent to say that the boundary-driven quantum spin chains cannot be ignored, nor avoided. Besides being phenomenologically justified in terms of the repeated interaction (RI) protocol \cite{KaPla}, these
boundary-driven quantum spin models can be experimentally realized \cite{real1, real2}. More importantly, in many real physical situations we are obligated to go beyond the models of systems weakly coupled to thermal baths, such as in computation and information processes using feedback control and in several  other problems of nonequilibrium statistical physics, see Ref.\cite{EspoPRX} and references there in.

The present work is addressed to the detailed investigation of the energy current and its components, heat and work, in the boundary-driven $XXZ$ spin chains. Besides the disclosure of fundamental features of the theme,
we are particularly motivated by the necessity of making clear the thermodynamical aspects of a nontrivial recent result \cite{Prapid2017}: the fact that boundary-driven, asymmetric (graded) $XXZ$ spin chains are genuine rectifiers of energy current. More interesting, in the absence of a magnetic field in the bulk of the chain, there is an one-way street for the energy current, i.e., the direction of the current is given by arrangements of asymmetries in the chain - the current
does not change as we invert the baths at the boundaries of these graded $XXZ$ chains. These findings announce promising applications in the control and manipulation of the energy current, and, in the present paper, we will make clear that there is no thermodynamical inconsistency in such results. Of course, a naive interpretation involving only heat in such phenomenon for the boundary-driven spin chain, i.e., if one considers the reservoirs only as heat baths at different temperatures and the energy current as heat between baths and system, then an incongruity appears.

To carry out our investigation, we first consider the RI scheme in order to
derive the related Lindblad master equation (LME) for the system, as well as the specific expressions for the currents of heat and work, exchanged between the system and the baths. After that, we obtain another expression for the energy current, now derived by means of a continuity equation for the energy and the LME, i.e., without using the repeated interaction scheme. This latter expression is usually found in the literature about transport in
the $XXZ$ chain. We make a comparison between the expressions, i.e., the one derived from the RI scheme and the other from the continuity equation and LME. It allows us to show the consistency of both approaches and to separate
heat and power coming from the baths to the system, even in the commonly used expression for the energy current (derived from the LME), in which such a distinction is far from being clear. The recognition of the work between baths and system allows us to understand the thermodynamical aspects of the transport process.

In the derivation of the formulas for heat and power by means of the RI scheme, we follow the procedure precisely and clearly described by Barra in Ref.\cite{FBarra}, where the idea of a distinction between heat and work in the
boundary-driven spin chains is proposed, and, besides the general approach, an example considering the
simpler case of the $XY$ model is presented. For completeness, in the Section II-B we extend some computations or arguments already presented there.

The rest of the paper is organized as follows. In section II, we describe the RI scheme. Then, in two subsections, we use the scheme to derive the LME and, soon after, to obtain the expressions for the heat and work exchanged between
 system and baths. In section III, we compare the usual expression for the energy current derived from the LME (and a continuity equation) with the expressions developed in Section II. The last section, Final Remarks, is
devoted to some further results and to the concluding remarks.

\section{The Repeated Interaction Scheme}

We describe now the RI scheme, which will lead us to the expressions for the boundary-driven LME, and for the heat and
work exchanged between system and baths. For the physical interest and importance of such an approach see, e.g., Ref.\cite{EspoPRX} and references there in.

The scheme is builded as follows. We take a system (in our particular case, a spin chain with time-independent Hamiltonian $H_{S}$), and couple it to two
baths, $H_{R}$ at the right boundary and $H_{L}$ at the left one. At $t=0$, or at the beginning of each time interval described below,  we assume that the system is decoupled from the baths, i.e.,
$$
\rho_{tot}(0) = \rho_{S}(0)\otimes\rho_{E}(0) ~,
$$
where $\rho$ denotes the density matrix; the index $E$ is for the environment, i.e., left and right baths. Then we couple the left and right baths to the
system and allow it to evolve up to a time $\tau$. After that, we take the partial trace over the first baths, couple the system again to a second copy of
baths, allow it to evolve from time $\tau$ to $2\tau$, take the partial trace and repeat the process indefinitely. For clearness, borrowing the notation from
Ref.\cite{FBarra}, we write the baths (left and right) as an infinite collection of copies acting in different intervals of time. That is, we write the
 Hamiltonian of the baths as $H_{r} = \sum_{n} H_{r}^{n}$, where $r$ is $L$ or $R$, and each $H^{n}$ interacts with the system for the time in the interval $t \in [(n-1)\tau, n\tau)$. The interaction between system and baths is also written as $V(t) = V^{n}$, $V^{n} = V^{n}_{L} + V^{n}_{R}$, with $n \in [1,2,\ldots]$. For the density matrices of the baths, we use the notation $\rho_{E} = \otimes_{n}\rho_{n}$, where, at the beginning of each time interval (i.e., immediately before the coupling between system and baths), we take
 \begin{eqnarray*}
 \rho_{n} &=& \omega_{\beta_{L}}(H^{n}_{L})\otimes  \omega_{\beta_{L}}(H^{n}_{L})~, \\
  \omega_{\beta_{r}} &=& e^{-\beta H_{r}}/Z_{r}~,
 \end{eqnarray*}
 i.e., $\omega_{\beta_{r}}$ is the the Boltzmann-Gibbs distribution for the bath.

 Thus, according to the described process, at the end of the n-$th$ step, we have
\begin{equation}
\rho_{S}(n\tau) = Tr_{n}\{ U_{n}[\rho_{S}((n-1)\tau)\otimes\rho_{n}]U_{n}^{\dagger}\} ~,
\end{equation}
where $Tr_{n}$ denotes the trace over the n-$th$ copy of the baths, and
$$
U^{n} = \exp[ -i\tau( H_{S} + H^{n}_{L} + H^{n}_{R} + V^{n}]
$$
is the time evolution in the correspondent time interval; we take $\hbar = 1$.

\subsection{From the RI scheme to the boundary driven LME}

Now, we take the previously described discrete mapping  and analyze its continuum limit $\tau \rightarrow 0$ in order to derive the LME.
We need to say that there exist different derivations for the LME without the use of the RI scheme. See, e.g., Refs.\cite{BP, GZ} and references there in.

It is convenient to specify our case of interest, i.e., the interactions to be treated in the present work. We take, for the Hamiltonian of the system,
\begin{eqnarray}
\mathcal{H} &=& \sum_{i=1}^{N-1}\left\{ \alpha_{i,i+1}\left( \sigma_{i}^{x}\sigma_{i+1}^{x} + \sigma_{i}^{y}\sigma_{i+1}^{y} \right) + \Delta_{i,i+1}\sigma_{i}^{z}\sigma_{i+1}^{z} \right\}\nonumber \\
 & & + \sum_{i=1}^{N} \frac{h_{i}}{2}\sigma_{i}^{z} ~, \label{hamiltonian}
\end{eqnarray}
and, for the left bath and the interaction system-bath,
$$
H_{L} = \frac{h_{L}}{2}\sigma_{L}^{z} ~~, ~~V_{L}  = \zeta_{L}\left(\sigma_{L}^{x}\sigma_{1}^{x} +
\sigma_{L}^{y}\sigma_{1}^{y}\right) ~,
$$
and similarly for the interactions in the right (R) side. Details about  $\zeta$, the strength of the system-bath interaction, are presented ahead. The recurrently used uniform $XXZ$ model is given by taking
$\alpha_{i,i+1} \equiv \alpha$ and $\Delta_{i,i+1} \equiv \Delta$.

To develop the formalism, we first write the time dynamical equation for $\rho_{S}$ in a power series in $\tau$, the time interval length,
\begin{eqnarray} \label{ps}
\lefteqn{\rho_{S}(n\tau) = Tr_{n}\{ U_{n}[\rho_{S}((n-1)\tau)\otimes\rho_{n}]U_{n}^{\dagger}\}} \nonumber \\
 &=& Tr_{n}\left[ e^{-iH_{T}\tau}\rho_{S}((n-1)\tau)\otimes\rho_{n} e^{+iH_{T}\tau} \right] \nonumber \\
 &=& Tr_{n}\left[ \rho -i\tau[H_{T},\rho] - \frac{\tau^{2}}{2}[H_{T},[H_{T},\rho]] + \ldots \right] ~.
 \end{eqnarray}
For simplicity, we denoted $\rho_{S}\otimes\rho_{n}$ by $\rho$ above. Then, we take the partial traces $Tr_{n}$. For the first term we obtain $Tr_{n}(\rho_{S}\otimes\rho_{n}) = \rho_{S}$. For the second one, due to our
specific choice of $V$, which does not involve $\sigma^{z}$, we have
$$
Tr_{n}\left( [H_{T}, \rho_{S}\otimes\rho_{n}] \right) = [H_{S},\rho_{S}] ~.
$$
To continue the analysis by taking the limit $\tau\rightarrow 0$, we note that, with the interaction time going to zero, we will eventually vanish the interaction between system and baths. In order to obtain a finite value
within such a procedure, we take $V$ properly increasing with $\tau$. We write $\zeta_{L} \equiv \sqrt{\lambda_{L}/\tau}$, and so,
$$
V_{L} = \sqrt{\frac{\lambda_{L}}{\tau}} \left( \sigma_{L}^{x}\sigma_{1}^{x} + \sigma_{L}^{y}\sigma_{1}^{y} \right) ~,
$$
and a similar expression follows for $V_{R}$. Turning to the Eq.(\ref{ps}) above, after some algebra we obtain
\begin{eqnarray*}
\lefteqn{\rho_{S}(n\tau) = \rho_{S}((n-1)\tau) - i\tau[H_{S},\rho_{S}((n-1)\tau)]~ +} \\
&&  \tau \left[ \mathcal{L}_{L}(\rho_{S}((n-1)\tau)) + \mathcal{L}_{R}(\rho_{S}((n-1)\tau)) \right] + \mathcal{O}(\tau^{>1}) ~,
\end{eqnarray*}
where
$$
\tau\mathcal{L}_{L}(\rho_{S}) = -\frac{\tau^{2}}{2} Tr_{n}[V_{L},[V_{L},\rho_{S}]] ~,
$$
and similarly for $\mathcal{L}_{R}(\rho_{S})$.

Hence, we write $\rho_{S}(n\tau) - \rho_{S}((n-1)\tau)$, divide by $\tau$, take the limit $\tau\rightarrow 0$, and obtain the LME. Namely, after a the scaling $\lambda_{r} \rightarrow \Gamma_{r}/4$,
\begin{eqnarray}
\dot{\rho}_{S} &=& -i[H_{S},\rho_{S}] + \mathcal{L}_{L}(\rho_{S}) +  \mathcal{L}_{R}(\rho_{S}) ~,\\
 \mathcal{L}_{L,R}(\rho_{S}) &=& \sum_{k=\pm} L_{k}\rho_{S}L^{\dagger}_{k} - \frac{1}{2}\{L^{\dagger}_{k}L_{k}, \rho_{S}\} ~,
\end{eqnarray}
where, for $\mathcal{L}_{L}$, we have
\begin{equation}
L_{\pm} = \sqrt{\frac{\Gamma_{L}}{2}(1 \pm f_{L})} \sigma_{1}^{\pm} ~\label{dissipator2},
\end{equation}
and similarly for $\mathcal{L}_{R}$, but with $\Gamma_{L}$, $\sigma_{1}^{\pm}$ and $f_{L}$ replaced by  $\Gamma_{R}$, $\sigma_{N}^{\pm}$ and $f_{R}$. In these expressions, we use the notation $\{\cdot,\cdot\}$ for the anticommutator; $\Gamma$ is
the coupling strength to the baths; $f_{L} = \langle \sigma_{L}^{z}\rangle$ and $f_{R} = \langle \sigma_{R}^{z}\rangle$ are the bath spin polarization at the boundaries, i.e., they give the driving strength;
$\sigma^{\pm}_{j} \equiv (\sigma^{x}_{j} + \sigma^{y}_{j})/2$ are the spin creation and annihilation operators. For simplicity, we take $\Gamma_{L} = \Gamma_{R} = \Gamma$.

In specific, written in terms of $\sigma_{1}^{\pm}$ and $\sigma_{N}^{\pm}$, the dissipator $\mathcal{L}(\rho_{s}) \equiv \mathcal{L}_{L}(\rho_{S}) +  \mathcal{L}_{R}(\rho_{S})$ in the LME becomes
\begin{widetext}
\begin{eqnarray}
\mathcal{L}(\rho_{S}) &=& \frac{\Gamma}{4}\left\{(1+f_{L})\left[2\sigma_{1}^{+}\rho_{S}\sigma_{1}^{-} - \left( \sigma_{1}^{-}\sigma_{1}^{+}\rho_{S}  + \rho_{S} \sigma_{1}^{-}\sigma_{1}^{+}\right)\right]\right.
                      + (1-f_{L})\left[2\sigma_{1}^{-}\rho_{S}\sigma_{1}^{+} - \left( \sigma_{1}^{+}\sigma_{1}^{-}\rho_{S}  + \rho_{S} \sigma_{1}^{+}\sigma_{1}^{-}\right)\right]  \nonumber\\
 &&  + (1+f_{R})\left[2\sigma_{N}^{+}\rho_{S}\sigma_{N}^{-} - \left( \sigma_{N}^{-}\sigma_{N}^{+}\rho_{S}  + \rho_{S} \sigma_{N}^{-}\sigma_{N}^{+}\right)\right]
                    \left.  + (1-f_{R})\left[2\sigma_{N}^{-}\rho_{S}\sigma_{N}^{+} - \left( \sigma_{N}^{+}\sigma_{N}^{-}\rho_{S}  + \rho_{S} \sigma_{N}^{+}\sigma_{N}^{-}\right)\right] \right\}.
\end{eqnarray}
\end{widetext}

A further note: for a comparison with other different works, it is convenient to use both notations $\lambda$ and $\Gamma$, where $\lambda_{L}=\Gamma_{L}/4$ as already presented (the same for the right side). It is also pertinent
to consider $B$ and $h$, where $B_{L} = h_{L}/2$, $B_{j} = h_{j}/2$ for $j$ a site in the bulk of the chain, etc.

\subsection{From the RI scheme to the expressions for heat and work}

Here, we follow the approach presented in Ref.(\cite{FBarra}). For clearness, in the first steps we repeat some calculations presented there for a general case; then we turn to the
computation of our specific case, namely, the $XXZ$ chain.

For a system with arbitrary strength coupling with the environment, the internal energy is better defined as
\begin{equation}
E(t) = Tr( \rho_{tot}(t)[H_{S}(t) + V(t)]) ~,
\end{equation}
where $Tr$ denotes the the full trace. And, by the first law of thermodynamics, its change is related to heat and work
$$
\Delta E(t) = W(t) + Q(t) ~,
$$
where $W(t)$ is the work performed on the system in the interval $[0,t]$, and it is defined as
\begin{equation}
W(t) = Tr( \rho_{tot}(t)H(t) - \rho_{tot}(0)H(0))~.
\end{equation}
The heat flow is given by $Q(t) = \sum_{r}Q_{r}(t)$, $r= R, L$, with
\begin{equation}
Q_{r}(t) = Tr( H_{r}\rho_{tot}(0) - H_{r}\rho_{tot}(t) )~,
\end{equation}
that is, the heat flow is given by minus the change in the energy of the r-$th$ bath.

To continue, within the repeated interaction scheme, we analyze each interval of time. Precisely, for $t \in [(n-1)\tau, n\tau)$, from the definitions above, we
have
$$
\Delta Q_{r} = Tr(H_{r}^{n}(\rho_{n} - \rho'_{n})~,
$$
where
\begin{eqnarray*}
\rho'_{n} &=& Tr_{S}(U_{n}\rho_{S}((n-1)\tau)\otimes \rho_{n}U_{n}^{\dagger}) ~,\\
U_{n} &=& e^{-i\tau(H_{S} +H^{n}_{L} + H^{n}_{R} + V^{n})} ~,
\end{eqnarray*}
i.e., $U_{n}$ is the time evolution in the interval $[(n-1)\tau,n\tau)$.

Now, we essentially repeat the procedure described in the derivation of the LME. We expand $U_{n}$ in powers of $\tau$:
writing $V = v/\sqrt{\tau}$, we have
$$
U_{n} = I - i\tau^{1/2}v - \tau(H_{0} +\frac{v^{2}}{2}) - \tau^{3/2}\frac{1}{2}\{H_{0}, v\} + \mathcal{O}(\tau^{2}) ~,
$$
where $H_{0} = H_{S} + H_{L} + H_{R}$. Then, for $\Delta Q_{r}$, considering that $Tr_{r}(v_{r}\omega_{\beta_{r}}) = 0$ and
$Tr_{r}[H_{r}, H_{0}] = 0$, we obtain
\begin{widetext}
\begin{eqnarray}
\Delta Q_{r} &=& - \tau Tr\left( \left(v_{r}H_{r}v_{r} - \frac{1}{2}\{v_{r}^{2},H_{r}\}\right)\rho_{S}((n-1)\tau)\otimes\omega_{\beta_{r}}\right) ~,\nonumber \\
\Rightarrow \dot{Q}_{r} &=& \lim_{\tau\rightarrow 0} \frac{\Delta Q_{r}}{\tau} =  - Tr\left( \left(v_{r}H_{r}v_{r} - \frac{1}{2}\{v_{r}^{2},H_{r}\}\right)\rho_{S}(t)\otimes\omega_{\beta_{r}}\right) ~.
\end{eqnarray}
\end{widetext}

After some
algebra, for our specific model, we get
\begin{eqnarray}
\dot{Q}_{L} &=& 2\lambda_{L}h_{L} \left[ M_{L} - Tr_{S}\left(\sigma_{1}^{z}\rho_{S}(t)\right)\right] \nonumber\\
&=& \Gamma_{L} B_{L} \left[ M_{L} - Tr_{S}\left(\sigma_{1}^{z}\rho_{S}(t)\right)\right] ~,\label{calor}
\end{eqnarray}
where
\begin{equation*}
M_{L} \equiv Tr_{L}\left(\sigma_{L}^{z}\omega_{\beta_{L}}\right) = -\tanh\left(\beta_{L}\frac{h_{L}}{2}\right) = -\tanh(\beta_{L}B_{L})~.
\end{equation*}

Now we develop the expression for the work.

Starting with the previous notation $V_{r}^{n}$ for $V_{r}(t)$, with $t \in [(n-1)\tau,n\tau)$, we have $\Delta W = \Delta W_{L} + \Delta W_{R}$ and
$\Delta W_{L} = Tr\left( [V^{n+1} - V^{n}]\rho_{tot}\right)$, where $\Delta W$ describes the work between the times $n\tau - \epsilon$ and
$n\tau + \epsilon$ (when we exchange the potential $V^{n}$ to $V^{n+1}$), say in the limit of $\epsilon\rightarrow 0$. As
$$
Tr_{r}\left(V_{r}^{n+1}\omega_{\beta_{r}}(H^{n+1})\right) =0~,
$$
 we stay with
$$
\Delta W_{r} = -Tr\left(V_{r}^{n}U_{n}\rho_{S}((n-1)\tau)\otimes\rho_{n} U_{n}^{\dagger}\right) ~.
$$
Now, as $V = v/\sqrt{\tau}$, in the expansion of $U_{n}$ in powers of $\tau$, we need to keep terms up to $\mathcal{O}(\tau^{3/2})$. After some algebra, we
obtain
\begin{widetext}
\begin{equation}
\dot{W}_{r} = lim_{\tau\rightarrow 0} \frac{\Delta W_{r}}{\tau} = Tr\left( \left(v_{r}(H_{S} + H_{r})v_{r} - \frac{1}{2}\{v_{r}^{2}, H_{S}+H_{r}\}\right)
\rho_{S}(t)\otimes\omega_{\beta_{r}}\right).
\end{equation}

Hence, performing the computation for the specific $H_{S}$ of the $XXZ$ chain, after some algebra we have
\begin{eqnarray}
\dot{W}_{L} &=& 2h_{1}\lambda_{L}\left[M_{L} - Tr_{S}(\sigma_{1}^{z}\rho_{S}(t))\right] -
2h_{L}\lambda_{L}\left[M_{L} - Tr_{S}(\sigma_{1}^{z}\rho_{S}(t))\right] \nonumber\\
& & - 2\lambda_{L} Tr_{S}\left( [\alpha(\sigma_{1}^{x}\sigma_{2}^{x} + \sigma_{1}^{y}\sigma_{2}^{y}) +
\Delta_{1,2}\sigma_{1}^{z}\sigma_{2}^{z}]\rho_{S}(t)\right) \nonumber \\
& & - 2\lambda_{L}\Delta_{1,2}Tr_{S}\left(\sigma_{1}^{z}\sigma_{2}^{z}\rho_{S}(t)\right) + 4\lambda_{L}\Delta_{1,2}M_{L}Tr_{S}\left(\sigma_{2}^{z}\rho_{S}(t)\right) ~.
\end{eqnarray}
\end{widetext}

Thus, adding the expressions for $\dot{W}_{L}$ and $\dot{Q}_{L}$, and the similar ones for the contact with the right bath, we obtain the energy rate
\begin{equation*}
\dot{E} = \dot{W}_{L} +\dot{Q}_{L} + \dot{W}_{R} +\dot{Q}_{R} ~.
\end{equation*}
A short comment: for the simpler case of a $XY$ model, when $\Delta \equiv 0$, and for the case $h_{1} = h_{L}$ (as well as $h_{N} = h_{R}$), the expressions
above for heat and power become those derived in Ref.\cite{FBarra}, see second line below Eq.(14) there in.

\section{Comparison between energy currents in the literature}

In the extensive literature about the $XXZ$ and related boundary driven quantum spin systems, we find several works devoted to the detailed analysis of
the spin and/or energy current. The expressions for the currents are usually derived from the LME, without any reference to the process, to the protocol which
leads to the LME \cite{Mendoza-A}. Then, the currents are analyzed in the stationary nonequilibrium state, reached as $t \rightarrow \infty$. The energy current, in particular, is treated as heat current in several works, without further justification or comments. Here, with the aim of elucidate any doubt and correct possible misinterpretations in the literature, we compare
the equations obtained via the RI scheme with those directly derived from the LME and make clear this point.

The derivation of the expression for the energy current by starting from the LME involves the use of a continuity equation for the energy flow. Precisely,
the Hamiltonian for the system is written as the sum of interparticle potentials,
\begin{eqnarray}
H &=&  \sum_{i=1}^{N-1} \varepsilon_{i,i+1} = \sum_{i=1}^{N-1} h_{i,i+1} + b_{i,i+1} ~,  \\
h_{i,i+1} &=& \alpha \left( \sigma_{i}^{x} \sigma_{i+1}^{x} + \sigma_{i}^{y}\sigma_{i+1}^{y} \right) + \Delta_{i,i+1} \sigma_{i}^{z} \sigma_{i+1}^{z} ~, \nonumber\\
b_{i,i+1} &=& \frac{1}{2} \left[ B_{i}\sigma_{i}^{z}(1+\delta_{i,1}) + B_{i+1}\sigma_{i+1}^{z}(1+\delta_{i+1,N}) \right]~,\nonumber
\end{eqnarray}
i.e., we split the Hamiltonian into the part related to the XXZ interaction and the part related to the external magnetic field. We take $\alpha_{i,i+1} = \alpha$ in the Hamiltonian above.
 And then, its time evolution (given by the LME) is investigated in association with a continuity equation
$$
\frac{d\langle \varepsilon_{i,i+1}\rangle}{dt} = - \left( \langle F_{i+1}\rangle - \langle F_{i}\rangle \right) ~,
$$
where $\langle \cdot\rangle$ is the average obtained by taking the trace over the system with the density matrix $\rho_{S}$; $\langle F_{i} \rangle$ is identified as the energy current at site $i$; and
$d\langle \varepsilon_{i,i+1}\rangle/dt = 0$ in the steady state.
Precisely, we have
\begin{eqnarray}
\lefteqn{\frac{d\langle \varepsilon_{i,i+1}\rangle}{dt} = \frac{d}{dt} \left( Tr(\rho\varepsilon_{i,i+1})\right) = Tr\left(\frac{d\rho}{dt}\varepsilon_{i,i+1}\right)} \nonumber\\
&=& -i Tr([H,\rho]\varepsilon_{i,i+1}) \nonumber\\
&& +~ Tr(\mathcal{L}_{L}(\rho)\varepsilon_{i,i+1}) + Tr(\mathcal{L}_{R}(\rho)\varepsilon_{i,i+1}) \nonumber\\
&=& i \langle[H,\varepsilon_{i,i+1}]\rangle + Tr(\mathcal{L}_{L}(\rho)\varepsilon_{1,2})\delta_{i,1} \nonumber\\
&& +~ Tr(\mathcal{L}_{R}(\rho)\varepsilon_{N-1,N})\delta_{i+1,N} \nonumber\\
&=& i\langle[\varepsilon_{i-1,i}~, \varepsilon_{i,i+1}]\rangle + i\langle[\varepsilon_{i+1,i+2}~, \varepsilon_{i,i+1}]\rangle \\
&& +~ Tr(\mathcal{L}_{L}(\rho)\varepsilon_{1,2})\delta_{i,1}  + Tr(\mathcal{L}_{R}(\rho)\varepsilon_{N-1,N})\delta_{i+1,N}~.\nonumber
\end{eqnarray}
From the expression above and the continuity equation, it is immediate the identification
\begin{eqnarray}
\langle F_{i} \rangle &\equiv & i\langle[\varepsilon_{i-1,i}~, \varepsilon_{i,i+1}]\rangle ~,~1<i<N ~,\nonumber \\
\langle F_{1} \rangle &\equiv & Tr(\mathcal{L}_{L}(\rho)\varepsilon_{1,2}) ~,\nonumber\\
\langle F_{N} \rangle &\equiv & -Tr(\mathcal{L}_{R}(\rho)\varepsilon_{N-1,N})~.
\end{eqnarray}
More details are presented, e.g., in Refs.\cite{Mendoza-A, SPL}.  Let us use the notation
$$
\langle F_{i}\rangle \equiv \langle F_{i}^{XXZ}\rangle + \langle F_{i}^{B}\rangle ~.
$$
Taking such an approach (in what follows, we consider the expressions in the
steady state),  in particular for $i=1$, i.e., for the site linked to the left bath, we obtain
\begin{eqnarray}
\lefteqn{\langle F_{1}^{XXZ}\rangle = \Gamma_{L}f_{L}\Delta_{1,2}\langle\sigma_{2}^{z}\rangle}  \\
&& -\frac{\Gamma_{L}}{2} \left( \langle \alpha(\sigma_{1}^{x}\sigma_{2}^{x} + \sigma_{1}^{y}\sigma_{2}^{y}) +
\Delta_{1,2}\sigma_{1}^{z}\sigma_{2}^{z}\rangle + \langle\Delta_{1,2}\sigma_{1}^{z}\sigma_{2}^{z}\rangle \right)  \nonumber \\
&& \langle F_{1}^{B}\rangle = \Gamma_{L}B_{1}\left(f_{L} - \langle\sigma_{1}^{z}\rangle \right) ~,
\end{eqnarray}
where $f_{L}$ is $M_{L}$ in the notation of Ref.\cite{FBarra}, $\Gamma_{L} = 4\lambda_{L}$ and $B_{1} = h_{1}/2$, as previously pointed out.

We can immediately verify that
\begin{equation}
\langle F_{1}\rangle \equiv \langle F_{1}^{XXZ}\rangle + \langle F_{1}^{B}\rangle = \dot{W}_{L} + \dot{Q}_{L} ~.
\end{equation}
In other words, the phenomenological, first principle derivation via the RI scheme, as proposed in Ref.\cite{FBarra}, as well as the usual derivation via continuity equation and LME, as
recurrently found in the literature, lead to the same result for the total energy current. However, only the RI scheme allows us to split the energy from the bath to the system (and vice-versa) into heat and work.

To further our analysis, we turn to the derivation of the expression for the energy current in the bulk of the chain, i.e., for the sites $i$ with $2 \leq i \leq N-1$. As described above, for $i$ in the bulk of the chain,
$$
\frac{d\langle \varepsilon_{i,i+1}\rangle}{dt}  = Tr\left(\frac{d\rho}{dt}\varepsilon_{i,i+1}\right) = -i Tr([H,\rho]\varepsilon_{i,i+1})~,
$$
that is, the dissipators do not directly appear in the equation, which is given only by changes in the density matrix, ruled by the commutation with the Hamiltonian.
Consequently, in several works, the authors name as heat current the energy current in this boundary-driven spin chains, see e.g. Ref.\cite{Mendoza-A}. It might be argued that it is acceptable for
the current in the bulk of the chain, but it is certainly incorrect for the energy current exchanged with the baths at the edges.

\section{Final Remarks}

For further comments and analysis, it is pertinent to consider the expression for the magnetization flow, i.e., for the spin current. Again, starting from the dynamics given by the LME and some continuity equations,
namely,
\begin{eqnarray}
\frac{ d\langle \sigma_{1}\rangle}{dt} &=& \langle J_{L} \rangle - \langle J_{1} \rangle ~, \nonumber \\
\frac{ d\langle \sigma_{i}\rangle}{dt} &=& \langle J_{i-1} \rangle - \langle J_{i} \rangle ~,  ~ 1< i < N ~, \nonumber \\
\frac{ d\langle \sigma_{N}\rangle}{dt} &=& \langle J_{N-1} \rangle - \langle J_{R} \rangle ~,
\end{eqnarray}
we obtain the expressions for the spin current. In particular, at the boundaries,
\begin{eqnarray}
\langle J_{L} \rangle &=& \gamma \left( M_{L} - \langle \sigma_{1}^{z} \rangle \right) ~,  \label{spinc} \nonumber \\
\langle J_{R} \rangle &=& -\gamma \left(M_{R} - \langle \sigma_{N}^{z} \rangle \right) ~.
\end{eqnarray}
From these previous equations, it follows that $\dot{Q}_{L} = B_{L}J_{L}$ and $\dot{Q}_{R} = -B_{R}J_{R}$, see Eq.(\ref{calor}).  And so, in the steady state, as $\langle J_{L}\rangle = \langle J_{R} \rangle = J$, we have
 $\dot{Q}_{L} + \dot{Q}_{R} = (B_{L} - B_{R})J$. That is, for these boundary-driven spin chains, in the nonequilibrium stationary state, we obviously have $\langle \dot{H}_{S} \rangle = \dot{Q} + \dot{W} = 0$, but
 $\dot{Q}$ does not necessarily vanish. Anyway, we do not have any thermodynamic dilemma. For example, let us consider the entropy production rate, defined as the difference between the time derivative of the von Neumann entropy and the  entropy flow
 $$
 \frac{d S}{dt} = -Tr_{S}\left( \mathcal{L}(\rho_{S}(t))) \ln\rho_{S}(t) \right) - \sum\beta_{r}\dot{Q}_{r} ~,
 $$
 where $\beta_{r} = 1/T_{r}$ (we take $K_{B}=1$). It is expected to be a nonnegative function. We analyze it in the steady state, where, from the expressions above, we obtain
\begin{equation*}
\frac{d S}{dt} = \left(\beta_{R} - \beta_{L}\right) J ~.
\end{equation*}
For the XXZ asymmetric chain, in which nontrivial behaviors for the energy current have been described \cite{SPL, Prapid2017}, it is very difficult to obtain the exact expression for the spin current $J$. But, for the simpler case of
a small chain of 3 sites, it is described in Ref.\cite{SPL} (see the appendix in there; numerical simulations in larger chains are also presented in this reference). It follows that, for the case of
$f = M_{L} = -\tanh\beta_{L}B_{L}$ and $M_{R} = -\tanh\beta_{R}B_{R} = -f$, we obtain $J$ proportional to $f$, for small $f$. See also some relations between $J$ and $f$, obtained from symmetries in the LME and described in
Ref.\cite{Prapid2017}. Thus, in such a situation, we have $\beta_{R}B_{R} = -  \beta_{L}B_{L}$ and $f \propto \beta_{R}B_{R}$. Consequently,
$$
\frac{d S}{dt} \propto (\beta_{R}B_{R} + \beta_{R}B_{R})(\beta_{R}B_{R}) \propto (\beta_{R}B_{R})^{2} \geq 0.
$$

As a further remark, we recall that in the  boundary-driven XXZ spin chain,  an archetypical model of nonequilibrium open quantum system,  we may find uncommon phenomena of transport, as previously stressed. In particular, for its
asymmetric (graded) version, we have the one-way street for the energy flow \cite{Prapid2017, SPL}, which means that the direction of the energy flow does not change as
we invert the baths at the boundaries. It happens if the magnetic field is zero in the bulk of the chain, i.e., $B_{i} = 0$ for $i= 1, \ldots,N$ (of course, the fields modeling the baths, $B_{L}$ and $B_{R}$, are
nonzero). If we consider a naive interpretation assuming  only heat in the whole process, i.e., an interpretation that assumes the reservoirs as single heat baths and the energy from baths to system as heat current, then we will find a conflict with the
 Clausius statement of the second law of thermodynamics. Now, aware of this differentiation between work and heat in the energy exchanged between the system and the baths at the boundaries, we know that the boundary-driven process brings work to the system, and so, we
 understand the plausibility  of such phenomenon. We can also see that the system (spin chain plus baths) may have different behaviors  according to the values of $B_{R}/B_{L}$ and $\beta_{R}/\beta_{L}$, i.e., it may act as a heater, a refrigerator, etc. Finally, we stress that no thermodynamical inconsistency seems to be present.

To conclude, we want to say that, in spite of the existence of substantial results in the study of the transport properties of interacting quantum chains, in particular, of the boundary-driven $XXZ$ model, see e.g.
Refs.\cite{Mendoza-A, SPL, Prapid2017, PopLi, Prosen2} and several references there in, we still have many open questions, and a more complete picture is desirable. With the present work, we hope to make clear some thermodynamical
imbroglio related to the energy current of the 1D $XXZ$ chain, and so, we expect to stimulate more research on the subject.






\vspace*{1 cm} {\bf Acknowledgments:} This work was partially supported by CNPq (Brazil).

\end{document}